\documentclass[11pt]{elsart}

\usepackage{graphicx}
\usepackage{psfrag}
\usepackage{epsfig}

\usepackage{amssymb}

\begin{document}

\begin{frontmatter}



\title{Exploring networks with traceroute-like probes: 
theory and simulations}


\author[label1]{Luca Dall'Asta},
\author[label1]{Ignacio Alvarez-Hamelin},
\author[label1]{Alain Barrat},
\author[label2]{Alexei V{\'a}zquez},
\author[label1,label3]{Alessandro Vespignani}

\address[label1]{Laboratoire de Physique Th\'eorique, B\^atiment 210,
Universit\'e de Paris-Sud, 91405 ORSAY Cedex France}
\address[label2]{Nieuwland Science Hall,
University of Notre Dame, Notre Dame, IN 46556, USA.}
\address[label3]{School of Informatics and Department of
Physics, University of Indiana, Bloomington, IN 47408, USA}

\begin{abstract}
Mapping the Internet generally consists in  sampling the network 
from a limited set of sources by using \texttt{traceroute}-like probes. 
This methodology, akin to the merging of different spanning trees to 
a set of destination, has been
argued to introduce uncontrolled sampling biases that might produce
statistical properties of the sampled graph which sharply differ from
the original ones\cite{crovella,clauset,delos}.
In this paper we explore these biases and provide a statistical analysis of
their origin. We derive an analytical approximation for the 
probability of edge and vertex detection that exploits the role of 
the number of sources and targets and allows us to relate the global
topological properties of the underlying network with the statistical
accuracy of the sampled graph. In particular, we find
that the edge and vertex detection probability depends on 
the {\em betweenness centrality} of each element. This allows us to 
show that shortest path routed
sampling provides a better characterization of underlying
graphs with broad distributions of connectivity. We complement the analytical 
discussion with a throughout numerical investigation of simulated
mapping strategies in network models with different topologies.
We show that sampled graphs provide a fair qualitative 
characterization of the statistical properties of the original 
networks in a fair range of different strategies and exploration 
parameters. Moreover, we characterize the level of redundancy and
completeness of the exploration process as a function of the topological
properties of the network. Finally, we study numerically how the fraction
of vertices and edges discovered in the sampled graph depends on
the particular deployements of probing sources. The results
might hint the steps toward more efficient mapping strategies.
\end{abstract}

\end{frontmatter}

\section{Introduction}

A significant research and technical challenge in the study 
of large information networks is related to the lack of highly 
accurate maps providing information on their basic topology.
This is mainly due to the dynamical nature 
of their structure and to the lack of any centralized control 
resulting in a self-organized growth and evolution of these systems.
A prototypical example of this situation is  faced in the case of 
the physical Internet. The topology of the Internet can be investigated
at different granularity levels such as the router and Autonomous
System (AS) level, with  the final aim of obtaining an abstract 
representation where the set of routers (ASs) and their physical 
connections (peering relations) are the vertices 
and edges of a graph, respectively. In the absence of accurate 
maps, researchers rely on a general strategy that consists 
in acquiring local views of the network from several 
vantage points and merging these views in order to get a presumably
accurate global map.
Local views are obtained by evaluating a
certain number of paths to different destinations by using specific 
tools such as \texttt{traceroute} or by the analysis of BGP tables. 
At first approximation these processes amount to the collection of
shortest paths from a source vertex to a set of target vertices, obtaining
a partial spanning tree of the network. The merging of several of
these views provides the map of the Internet from which the statistical
properties of the network are evaluated. 

By using this strategy, a number of research groups have generated 
maps of the Internet~\cite{nlanr,caida,asdata,scan,lucent}, that have 
been used for the statistical characterization of the network
properties. Defining $\mathcal{G}=(V,E)$
as the sampled graph of the Internet with $N=|V|$ vertices and $|E|$ edges, 
it is quite intuitive that the Internet is a {\em sparse} graph
in which the number of edges is much lower than in a complete graph;
i.e. $|E| \ll N(N-1)/2$. Equally important is the fact that the average
distance, measured as the shortest path, between vertices is very
small. This is the so called {\em small-world} property, that is
essential for the efficient functioning of the network.  
Most surprising is the evidence of a skewed and  heavy-tailed behavior  
for the probability that any vertex in the 
graph has degree $k$ defined as the number of edges linking each 
vertex to its neighbors. In particular, in several instances, the degree 
distribution appears to be  approximated by $P(k)\sim k^{-\gamma}$ 
with $2\le \gamma\le 2.5$ \cite{faloutsos}. Evidence
for the heavy-tailed behavior of the degree distribution has been 
collected in several other studies at the router and AS level 
\cite{mercator,broido,calda,us02,chen02} and have generated a large
activity in the field of network modeling and characterization 
\cite{brite,inet,mdbook,baldi,psvbook}.

While \texttt{traceroute}-driven strategies are very flexible and 
can be feasible for extensive use, the obtained maps are undoubtedly
incomplete. Along with technical problems such as the instability of
paths  between routers and  interface resolutions \cite{burch99},
typical mapping projects are run from relatively small sets of sources
whose combined  views are missing a considerable number of edges
and vertices \cite{chen02,willinger}. In particular, the various
spanning trees are specially missing the lateral connectivity of 
targets and sample more frequently vertices and links which are closer 
to each source, introducing spurious effects that might seriously
compromise the statistical accuracy of the sampled graph.
These {\em sampling biases}
have been explored in numerical experiments of synthetic graphs 
generated by different algorithms\cite{crovella,clauset,delos}. 
Very interestingly, it has been shown that apparent degree 
distributions with heavy-tails may be 
observed even from homogeneous topologies such as in the classic 
Erd{\"o}s-R{\'e}nyi graph model\cite{crovella,clauset}. 
These studies thus point out that the evidence obtained from the 
analysis of the Internet sampled graphs might be insufficient to draw
conclusions on the topology of the actual Internet network. 

In this work we tackle this problem by performing a mean-field statistical
analysis and extensive numerical study of shortest path routed sampling,
considered as the first approximation to \texttt{traceroute}-sampling
(see section \ref{sec:traceroute}), in different networks models.  We derive
in section \ref{stat} an approximate
expression for the probability of edges and vertices to be detected that
exploits the dependence upon the number of sources, targets and the
topological properties of the networks. The expression shows the dependency
of the efficiency of the mapping process upon the number of sources, targets
and the topological properties of the network.  Moreover, the analytical study
provides a general understanding of which kind of topologies yields the most
accurate sampling. In particular, we show that the map accuracy depends on
the underlying network {\em betweenness centrality} distribution; the heavier
the tail the higher the statistical accuracy of the sampled graph.

We substantiate our analytical finding with a throughout exploration of maps
obtained varying the number of source-target pairs on networks models 
with different topological properties. In particular, we consider
networks with degree distribution with poissonian, Weibull and
power-law behavior (see section \ref{models}).
According to the theoretical analysis, both the total number of probes
deployed and the topological properties seem to play a primary role in the
understanding of the level of the efficiency reached by the mapping process.
As a measure of the efficiency of the mapping in different network topologies,
we study the fractions of discovered vertices and edges as a function of the
degree (section \ref{efficiency}), stressing the agreement with the
theoretical predictions. Other interesting quantities such as 
transit frequency and traffic entropy, are introduced in the study
of the discovery process, with the aim of providing a complete framework for
the study of sampling redundancy (section
\ref{redundancy}). Furthermore we focus on the study of the degree
distributions obtained in the sampled graph and their resemblance to
the original ones (see Section
\ref{sec:pk}). Our results show that
single source mapping processes face serious limitations in that also the
targeting of the whole network results in a very partial discovery of its
connectivity. On the contrary, the use of multiple sources promptly leads to
obtained maps fairly consistent with the original sample, where the
statistical degree distributions are qualitatively discriminated also
at relatively low values of target density. A detailed discussion of 
the behavior of the degree distribution as a function of targets and
sources is provided for sampled graphs with different topologies and compared
with the insight obtained by analytical means.

In section \ref{sec:opt}, we also inspect quantitatively 
the portion of discovered network in different mapping strategies for
the deployment of sources that however impose the same density 
of probes to the network; i.e. having the same probing load. 
We find the presence of a region of low efficiency
(less vertices and edges discovered) depending on the relative proportion of
sources and targets. This low efficiency region however corresponds 
to the optimal estimation of the network average degree.
This finding calls for a
``trade-off'' between the accuracy in the observation of different quantities
and hints to possible optimization procedures in the
\texttt{traceroute}-driven mapping of large networks.

\section{Related work}

In this section, we shortly review some recent works devoted to the sampling
of graphs by shortest path probing procedures.
Lakhina et al. \cite{crovella} have shown that biases can seriously affect the
estimation of degree distributions. In particular, power-law
like distributions can be observed for subgraphs of Erd\"os-R\'enyi random
graphs when the subgraph is the product of a \texttt{traceroute} exploration
with relatively few sources and destinations. They discuss the origin of these
biases and the effect of the distance between source and target in the mapping
process. In a recent work \cite{clauset}, Clauset and Moore have given
analytical foundations to the numerical work of Lakhina et al.
\cite{crovella}.  They have modeled the single source probing to all possible
destinations using differential equations.  For an Erd\"os-Renyi random graph
with average degree $\overline{k}$, they have found that the connectivity
distribution of the obtained spanning tree displays a power-law behavior
$k^{-1}$, with an exponential cut-off setting in at a characteristic degree
$k_c\sim\overline{k}$.

In a slightly different context, Petermann and De Los Rios have studied a
\texttt{traceroute}-like procedure on various examples of scale-free graphs
\cite{delos},
showing that, in the case of a single source, power-law distributions with
underestimated exponents are obtained. Analytical estimates of the measured
exponents as a function of the true ones were also derived.  
Finally, a recent preprint by Guillaume and Latapy \cite{latapy} reports about
the shortest-paths explorations of synthetic graphs, focusing on the
comparison between properties of the resulting sampled graph with those of the
original network.  The proportion of discovered vertices and edges in the graph
as a function of the number of sources and targets gives also hints for an
optimization of the exploration process.  

All these pieces of work make clear
the relevance of determining up to which extent the topological properties
observed in sampled graphs are representative of that of the real networks.

\section{A theoretical model for \texttt{traceroute}-like processes}
\label{sec:traceroute}

\begin{figure}[thb]
\begin{center}
\includegraphics[width=5cm]{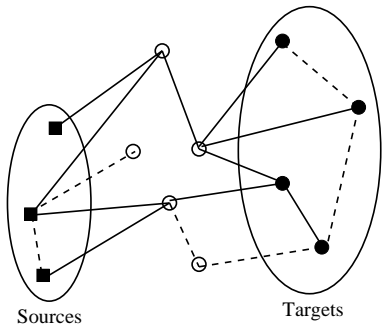}
\end{center}
\caption{Illustration of the \texttt{traceroute}-like procedure. 
Shortest paths between the set of sources and the set of destination 
targets are
discovered (shown in full lines) while other edges are not found
(dashed lines). Note that not all shortest paths are found since the
``Unique Shortest Path'' procedure is used. }
\label{fig:1}
\end{figure}

In a typical \texttt{traceroute} study, a set of active sources deployed in
the network sends \texttt{traceroute} probes to a set of destination vertices.
Each probe collects information on all the vertices and edges traversed along the
path connecting the source to the destination, allowing the discovery of the
network \cite{burch99}.  By merging the information collected on each path it
is then possible to reconstruct a partial map of the network
(Fig.\ref{fig:1}).  More in detail, the edges and the vertices discovered by each
probe will depend on the ``path selection criterium'' used to decide the path
between a pair of vertices. In the real Internet, many factors, including
commercial agreement, traffic congestion and administrative routing policies,
contribute to determine the actual path, causing it to differ even
considerably from the shortest path.  Despite these local, often unpredictable
path distortions or inflations, a reasonable first approximation of the route
traversed by \texttt{traceroute}-like probes is the shortest path between the
two vertices.  This assumption, however, is not sufficient for a proper
definition of a \texttt{traceroute} model in that equivalent shortest paths
between two vertices may exist.  In the presence of a degeneracy of shortest
paths we must therefore specify the path selection criterium by providing a
resolution algorithm for the selection of shortest paths.

For the sake of simplicity we can define three selection
mechanisms defining different ideal-paths  
that may account for some of the features encountered in Internet
discovery:
\begin{itemize}
\item Unique Shortest Path (USP) probe. In this case the shortest
path route selected between a vertex $i$ and the destination target $T$ 
is always the same independently of the source $S$ (the path being
initially chosen at random among all the equivalent ones). 

\item Random Shortest Path (RSP) probe. The shortest path between
any source-destination pair is chosen randomly among the set of
equivalent shortest paths. This might mimic different peering
agreements that make independent the paths among couples of vertices.

\item All Shortest Paths (ASP) probe. The selection criterium discovers all
  the equivalent shortest paths between source-destination pairs. This might
  happen in the case of probing repeated in time (long time exploration), so
  that back-up paths and equivalent paths are discovered in different runs.
\end{itemize}

We will generically call $\mathcal{M}$-path the path found using one of these
measurement or path selection mechanism.  Actual \texttt{traceroute} probes
contain a mixture of the three mechanisms defined above. We do not attempt,
however, to account for all the subtleties that real studies encounters, i.e.
IP routing, BGP policies, interface resolutions and many others.  In fact, in
the real mapping process, many effective heuristic strategies are commonly
applied to improve the reliability and the performances of the sampling.  For
instance, the interface resolution is well achieved by the {\em iffinder}
algorithm proposed by Broido and Claffy \cite{broido}. However, we will see
that the different path selection criteria (p.s.c.) have only little influence
on the general picture emerging from our results.  Moreover, the USP procedure
clearly represents the worst case scenario since, among the three different
methods, it yields the minimum number of discoveries. For this reason, if not
otherwise specified, we will report the USP data to illustrate the general
features of our synthetic exploration.  The interest of this analysis resides
properly in the choice of working in the most pessimistic case, being aware
that path inflations should actually provide a more pervasive sampling of the
real network.

More formally, the experimental setup for our simulated \texttt{traceroute}
mapping is the following. Let $G=(V,E)$ be a sparse undirected graph with
vertices (vertices) $V=\{1,2,\cdots,N\}$ and edges (links) $E$. Then let us
define the sets of vertices $\mathcal{ S}=\{i_1,i_2,\cdots,i_{N_S}\}$ and
$\mathcal{ T}=\{j_1,j_2,\cdots,j_{N_T}\}$ specifying the random placement of
$N_S$ sources and $N_T$ destination targets.  For each ensemble of
source-target pairs $\Omega=\{ \mathcal{S}, \mathcal{T} \}$, we compute with
our p.s.c. the paths connecting each source-target pair. The sampled graph
$\mathcal{G}=(V^*,E^*)$ is defined as the set of vertices $V^*$ (with
$N^*=|V^*|$) and edges $E^*$ induced by considering the union of all the
$\mathcal{M}$-paths connecting the source-target pairs.  The sampled graph is
thus analogous to the maps obtained from real \texttt{traceroute} sampling of
the Internet.

In our study the  parameters of interest are the density $\rho_T=N_T/N$
and $\rho_S=N_S/N$ of targets and sources.
In general, \texttt{traceroute}-driven studies run from a relatively
small number of sources to a much larger set of destinations.
For this reason, in many cases it is appropriate to work with 
the density of targets $\rho_T$ while still considering 
$N_S$ instead of the corresponding density. Indeed, it is 
clear that while $100$ targets may represent a fair probing of a 
network composed by $500$ vertices, this number would be clearly 
inadequate in a network of $10^6$ vertices. On the contrary, 
the density of targets $\rho_T$ allows us to compare mapping 
processes on networks with different sizes by
defining an intrinsic percentage of targeted vertices.
In many cases, as we will see in the next sections, an appropriate 
quantity representing the level of sampling of the networks is  
$\epsilon= \frac{N_S N_T} {N}$,
that measures the density of probes imposed to the system. In real situations
it represents the density of \texttt{traceroute} probes in the network and
therefore a measure of the load provided to the network by the measuring
infrastructure.

In the following, our aim is to evaluate to which extent the
statistical properties of the sampled graph $\mathcal{G}$ 
depend on the parameters of our experimental setup and are 
representative of the properties of the underlying graph $G$.

\section{Mean-field theory of simulated mapping process}
\label{stat}

We begin our study by presenting a mean-field statistical
analysis of the simulated \texttt{traceroute} mapping. Our aim is to
provide a statistical estimate for the probability of edge and vertex
detection as a function of $N_S$, $N_T$ and the topology of the
underlying graph. 

Let us define the quantity $\sigma_{i,j}^{(l,m)}$ that takes the value $1$ 
if the edge $(i,j)$ belongs to the selected $\mathcal{M}$-path 
between vertices $l$ and $m$, and $0$ otherwise. 
The indicator function
that a given edge $(i,j)$ will be discovered and belongs to
the sampled graph is given by 
\begin{equation}
\pi_{i,j}=1-
\prod_{l\neq m}\left(
1 - 
\sum_{s=1}^{N_S} \delta_{l,i_s} 
\sum_{t=1}^{N_T} \delta_{m,j_t} 
\sigma_{i,j}^{(l,m)}
\right), 
\end{equation}
where $\delta_{i,j}$ is the Kronecker symbol and selects only vertices
belonging to the set of sources or targets.
In the case of a given set $\Omega=\{ \mathcal{S}, \mathcal{T} \}$,
the above function is simply $\pi_{i,j}=1$ if the edge $(i,j)$ 
belongs to at least one of the $\mathcal{M}$-paths connecting the 
source-target pairs, and $0$ otherwise.
While the above exact expression does not lead us too far in the
understanding of the discovery probabilities, it is interesting to
look at the process on a statistical ground by studying the average over 
all possible realizations of the set $\Omega=\{ \mathcal{S}, \mathcal{T} \}$.
By definition we have that 
\begin{equation}
\left\langle \sum_{t=1}^{N_T} \delta_{i,j_t} \right\rangle
= \rho_T~~~\mbox{and}~~
\left\langle \sum_{s=1}^{N_S} \delta_{i,i_s} \right\rangle
= \rho_S,
\label{aver}
\end{equation}
where $\left\langle\cdots\right\rangle$ identifies the average over all
possible deployment of sources and targets $\Omega$.  These equalities simply
state that each vertex $i$ has, on average, a probability to be a source or a
target that is proportional to their respective densities.  In the following,
we will make use of an uncorrelation assumption that yields an explicit
approximation for the discovery probability. The assumption consists in
neglecting correlations originated by the position of sources and targets on
the discovery probability by different paths.  While this assumption does not
provide an exact treatment for the problem it generally conveys a qualitative
understanding of the statistical properties of the system.  In this
approximation, the average discovery probability of an edge is
\begin{eqnarray}
\nonumber&&\left\langle\pi_{i,j}\right\rangle=1-
\left\langle\prod_{l\neq m}\left(
1 - 
\sum_{s=1}^{N_S} \delta_{l,i_s} 
\sum_{t=1}^{N_T} \delta_{m,j_t} 
\sigma_{i,j}^{(l,m)}
\right)\right\rangle\\
&&~~~~~~~~~~\simeq 1-\prod_{l\neq m}\left(1 -
\rho_T\rho_S \left\langle\sigma_{i,j}^{(l,m)}\right\rangle\right),
\label{meanedge}
\end{eqnarray}
where in the last term we take advantage of neglecting correlations by
replacing the average of the product of variables with the product of the
averages and using Eq.~(\ref{aver}). This expression simply states that each
possible source-target pair weights in the average with the product of the
probability that the end vertices are a source and a target; the discovery
probability is thus obtained by considering the edge in an average effective
media ({\em mean-field}) of sources and targets homogeneously distributed in
the network. This approach is indeed akin to mean-field methods customarily
used in the study of many particle systems where each particle is considered
in an effective average medium defined by the uncorrelated averages of
quantities.  The realization average of
$\left\langle\sigma_{i,j}^{(l,m)}\right\rangle$ is very simple in the
uncorrelated picture, depending only of the kind of the probing model. In the
case of the ASP probing, $\left\langle\sigma_{i,j}^{(l,m)} \right\rangle$ is
just one if $(i,j)$ belongs to one of the shortest paths between $l$ and $m$,
and $0$ otherwise.  In the case of the USP and the RSP, on the contrary, only
one path among all the equivalent ones is chosen. If we denote by
$\sigma^{(l,m)}$ the number of shortest paths between vertices $l$ and $m$,
and by $x_{i,j}^{(l,m)}$ the number of these paths passing through the edge
$(i,j)$, the probability that the \texttt{traceroute} model chooses a path
going through the edge $(i,j)$ between $l$ and $m$ is
$\left\langle\sigma_{i,j}^{(l,m)}\right\rangle$ = $x^{(l,m)}_{i,j} /
\sigma^{(l,m)}$.

The standard situation we consider is the one in which 
$\rho_T\rho_S\ll 1$ and since 
$\left\langle\sigma_{i,j}^{(l,m)}\right\rangle\leq 1 $, we have  
\begin{equation}
\prod_{l\neq m}
\left(1 -\rho_T\rho_S \left\langle\sigma_{i,j}^{(l,m)}
\right\rangle\right)\simeq
\prod_{l\neq m}\exp \left(-\rho_T\rho_S
\left\langle\sigma_{i,j}^{(l,m)}\right\rangle\right),
\end{equation}
that inserted in Eq.(\ref{meanedge}) yields 
\begin{equation}
\nonumber\left\langle\pi_{i,j}\right\rangle \simeq 1-
\prod_{l\neq m} \left(\exp \left(- \rho_T\rho_S
\left\langle\sigma_{i,j}^{(l,m)}\right\rangle\right)\right)
= 1-\exp
\left(-\rho_T\rho_S b_{ij}
\right),
\label{edgedisc}
\end{equation}
where $b_{ij}=\sum_{l\neq m}\left\langle\sigma_{i,j}^{(l,m)}\right\rangle$. In
the case of the USP and RSP probing, the quantity $b_{ij}$ is by definition
the edge {\em betweenness centrality} $\sum_{l \neq m} x^{(l,m)}_{i,j} /
\sigma^{(l,m)}$ \cite{freeman77,brandes}, sometimes also refereed to as
``load''~\cite{goh01} (In the case of ASP probing, it is a closely related
quantity).  Indeed the vertex or edge betweenness is defined as the total
number of shortest paths among pairs of vertices in the network that pass
through a vertex or an edge, respectively. If there are multiple shortest
paths between a pair of vertices, the path contributes to the betweenness with
the corresponding relative weight.  The {\em betweenness} gives a
measure of the
amount of all-to-all traffic that goes through an edge or vertex, if the
shortest path is used as the metric defining the optimal path between pairs of
vertices, and it can be considered as a non-local measure of the
\textit{centrality} of an edge or vertex in the graph.

The edge betweenness assumes values between $2$ and $N(N-1)$ and the
discovery probability of the edge will therefore depend strongly on
its betweenness. In particular, for edges with minimum 
betweenness \mbox{$b_{ij}=2$} we have 
$\left\langle\pi_{i,j}\right\rangle \simeq 2\rho_T\rho_S$,
that recovers the probability that the two end vertices of the edge
are chosen as source and target. This implies that if the densities 
of sources and targets are small but finite in the limit of very large
$N$, all the edges in the underlying graph have an appreciable
probability to be discovered. Moreover, for edges
with high betweenness the discovery probability approaches one. A fair
sampling of the network is thus expected.

In most realistic samplings, however, we face a very different situation.
While it is reasonable to consider $\rho_T$ a small but finite value, the
number of sources is not extensive ($N_S\sim \mathcal{O}(1)$) and their
density tends to zero as $N^{-1}$. In this case it is more convenient to
express the edge discovery probability as
\begin{equation}
\left\langle\pi_{i,j}\right\rangle \simeq 
 1-\exp\left(-\epsilon \widetilde{b_{ij}}
\right),
\label{edgesample}
\end{equation}
where $\epsilon=\rho_T N_S$ is the density of probes imposed to the system and
the rescaled betweenness $\widetilde{b_{ij}}=N^{-1}b_{ij}$ is now limited in
the interval $[2N^{-1},N-1]$. In the limit of large networks $N\to\infty$ it
is clear that edges with low betweenness have
$\left\langle\pi_{i,j}\right\rangle \sim \mathcal{O}(N^{-1})$, for any finite
value of $\epsilon$. This readily implies that in real situations the
discovery process is generally not complete, a large part of low betweenness
edges being not discovered, and that the network sampling is made
progressively more accurate by increasing the density of probes $\epsilon$.

A similar analysis can be performed for the discovery probability
$\pi_{i}$ of a
vertex $i$. For each source-target set $\Omega$ we have that 
\begin{equation}
\pi_{i}=1-\left(
1-\sum_{s=1}^{N_S} \delta_{i,i_s}- \sum_{t=1}^{N_T} \delta_{i,j_t}
\right)
\prod_{l\neq m\neq i}\left(
1 - \sum_{s=1}^{N_S} \delta_{l,i_s} 
\sum_{t=1}^{N_T} \delta_{m,j_t} \sigma_{i}^{(l,m)}
\right).
\end{equation}
where $\sigma_{i}^{(l,m)}=1$ if the vertex $i$ belongs to the
$\mathcal{M}$-path between vertices $l$ and $m$, and $0$ otherwise.  This time it
has been considered that each vertex is discovered with probability one also
if it is in the set of sources and targets.  The second term on the right hand
side therefore expresses the fact that the vertex $i$ does not belong to the
set of sources and targets and it is not discovered by any $\mathcal{M}$-path
between source-target pairs.  By using the same {\em mean-field} approximation
as previously, the average vertex discovery probability reads as
\begin{equation}
\left\langle\pi_{i}\right\rangle \simeq 1 - (1-\rho_S-\rho_T)
\prod_{l\neq m\neq i}\left(1 -\rho_T\rho_S
\left\langle\sigma_{i}^{(l,m)}\right\rangle\right). 
\end{equation}
As for the case of the edge discovery probability, the average considers all
possible source-target pairs weighted with probability $\rho_T\rho_S$. In the
ASP model, the average $\left\langle\sigma_{i}^{(l,m)}\right\rangle$ is $1$ if
$i$ belongs to one of the shortest paths between $l$ and $m$, and $0$
otherwise.  For the USP and RSP models,
$\left\langle\sigma_{i}^{(l,m)}\right\rangle$ = $x^{(l,m)}_{i} /
\sigma^{(l,m)}$ where $x^{(l,m)}_{i}$ is the number of shortest paths between
$l$ and $m$ going through $i$.  If $\rho_T\rho_S\ll 1$, by using the same
approximations used for Eq.(\ref{edgedisc}) we obtain
\begin{equation}
\left\langle\pi_{i}\right\rangle\simeq 1 - (1-\rho_S-\rho_T)
\exp \left(-\rho_T\rho_S b_{i}
\right),
\end{equation}
where $b_{i}=\sum_{l\neq m\neq
i}\left\langle\sigma_{i}^{(l,m)}\right\rangle$. For the USP and RSP cases, 
$b_{i}=\sum_{l\neq m\neq i} x^{(l,m)}_{i} / \sigma^{(l,m)}$ is the vertex 
betweenness centrality, that is limited in the interval $[0,N(N-1)]$
\cite{freeman77,brandes,goh01}.
The betweenness value $b_i=0$ holds for the leafs of the graph,
i.e. vertices with a single edge, for
which we recover
$\left\langle\pi_{i}\right\rangle\simeq\rho_S+\rho_T$.
Indeed, this kind of vertices are dangling ends discovered only if
they are either a source or target themselves.

As discussed before, the most usual setup corresponds to a density 
$\rho_S\sim\mathcal{O}(N^{-1})$ and in the large $N$ limit we can
conveniently write 
\begin{equation}
\left\langle\pi_{i}\right\rangle\simeq 1 - (1-\rho_T)
\exp \left(-\epsilon\widetilde{b_{i}}
\right),
\label{vertexsample}
\end{equation}
where we have neglected terms of order $\mathcal{O}(N^{-1})$ and the rescaled
betweenness $\widetilde{b_{i}}=N^{-1}b_{i}$ is now defined in the interval
$[0,N-1]$.  This expression points out that the probability of vertex
discovery is favored by the deployment of a finite density of targets that
defines its lower bound.

We can also provide a simple approximation for the effective
average degree $\left\langle k_i^*\right\rangle$ of the vertex $i$ 
discovered by our sampling process. Each edge departing from the vertex 
will contribute proportionally to its discovery probability, yielding   
\begin{equation}
\left\langle k_i^*\right\rangle= 
\sum_j \left( 1 - \exp \left(-\epsilon\widetilde{b_{ij}}\right) \right)
\simeq \epsilon\sum_j\widetilde{b_{ij}}. 
\end{equation}
The final expression is obtained for  edges with 
$\epsilon\widetilde{b_{ij}}\ll 1$. Since
the sum over all neighbors of the edge betweenness is
simply related to the vertex betweenness as 
$\sum_j{b_{ij}}= 2(b_{i}+ N-1)$, where the factor $2$ considers that
each vertex path traverses two edges and the term $N-1$ accounts for
all the edge paths for which the vertex is an endpoint, this finally
yields 
\begin{equation}
\left\langle k_i^*\right\rangle\simeq
2\epsilon + 2\epsilon\widetilde{b_{i}}.
\label{degdisc}
\end{equation}

The present analysis shows that the measured quantities and statistical
properties of the sampled graph strongly depend on the parameters of the
experimental setup and the topology of the underlying graph. The latter
dependence is exploited by the key role played by edge and vertex betweenness
in the expressions characterizing the graph discovery. The betweenness is a
nonlocal topological quantity whose properties change considerably depending
on the kind of graph considered. This allows an intuitive understanding of the
fact that graphs with diverse topological properties deliver different answer
to sampling experiments.

\section{Definition of the graph models}
\label{models}

In  order
to investigate numerically the \texttt{traceroute}-like exploration process,
we have deliberately chosen simple models endowed with very well-defined
topological properties, so as to give a clear result on which kind of
topologies are related to good sampling performances and vice-versa. Starting
from this first investigation, further studies could deal with more
realistic models as those created using Internet topology generators
\cite{inet,brite}.

Let us consider sparse undirected graphs denoted by $G=(V,E)$ where the
topological properties of a graph are fully encoded in its adjacency matrix
$a_{ij}$, whose elements are $1$ if the edge $(i,j)$ exists, and $0$
otherwise.  In particular, we will consider two main classes of graphs: {\em
  i)} {\em Homogeneous graphs} in which the degree
distribution $P(k)$ has small fluctuations and a well defined average
degree; {\em ii)} {\em
  Heterogeneous graphs} for which $P(k)$ is a broad distribution with 
heavy-tail and large fluctuations.  In this context, the 
{\em homogeneity} refers to the existence of a meaningful
characteristic average
degree that represents the typical value in the graph. For instance, 
in graphs with poissonian-like degree distribution a vast majority 
of vertices has degree
close to the average value and deviations from the average are exponentially
small in number. On the contrary, graphs with heavy-tailed degree
distribution are characterized by a strong heterogeneity encoded in
the presence of very large fluctuations and degree values varying over 
a wide range of magnitudes.

\subsection{Models}

The most widely known model for homogeneous graphs is given by the classical
Erd\"os-R\'enyi (ER) model \cite{er}: in such random graphs $G_{N,p}$ of $N$
vertices, each edge is present in $E$ independently with probability $p$.  The
expected number of edges is therefore $|E|=pN(N-1)/2$. In order to have sparse
graphs one thus needs to have $p$ of order $1/N$, since the average degree is
$p(N-1)$.  Erd\"os-R\'enyi graphs are typical examples of homogeneous graphs,
with degree distribution following a Poisson law.  Since $G_{N,p}$ can consist
of more than one connected component, we consider only the largest of these
components.

In opposition to the previous case, heterogeneous graphs are 
characterized by connectivity distributions spanning
various orders of magnitude, with a heavy-tail at large $k$.
In the literature, different definitions of heavy-tailed like 
distributions exist.
While we do not want to enter the detailed definition of
heavy-tailed distribution we 
have considered two classes of such distributions: (i) {\em
scale-free} or Pareto 
distributions of the form $P(k) \sim k^{-\gamma}$ (RSF), and (ii) Weibull
distributions (WEI)
$P(k)=(a/c) (k/c)^{a-1} \exp(-(k/c)^a)$. 
The scale-free distribution has a diverging second moment and
therefore virtually unbounded fluctuations, limited only by eventual
size-cut-off. The Weibull distribution has a coefficient of variation 
larger than the one of  exponential distributions but is not
power-law tailed. This distribution is akin to power-law
distribution truncated by an exponential cut-off which are often
encountered in the analysis of scale-free systems in the real
world. Indeed, a truncation of the power-law behavior is generally due
to finite-size effects and other physical constraints.
Both forms have been proposed as representing the
topological properties of the Internet \cite{broido}. 
In both cases, we have generated the corresponding  random graphs by 
using the algorithm proposed by
Molloy and Reed \cite{molloy}.  It consists in assigning to the vertices of the
graph a fixed sequence of degrees $\{{k}_{i}\}$, \ $i = 1, \dots, N$,
chosen at random from the desired degree distribution $P(k)$, and with the
additional constraint that the sum $\sum_{i} {k}_{i}$ must be even.
Then, the vertices are connected by $\sum_{i} {k}_{i} / 2$ edges,
respecting the assigned degrees and avoiding self- and multiple-connections.
The parameters used are $a=0.25$ and $c=0.6$ for the Weibull distribution, and
$\gamma=2.3$ for the RSF case.
The main properties of the various graphs are summarized in table \ref{table}.
In all numerical studies we have used networks of $N=10^4$ vertices.  
It is noteworthy that the maximum value of the degree ($k_{max}$) is
of the same order as the average for homogeneous graphs, but much 
larger for heterogenous ones.

\begin{table}[b]
\caption{\small 
Main characteristics of the graphs used in the numerical exploration.
}
\begin{center}
\begin{tabular}{|c|c|c|c|c|c|c|}\hline
  & ER & ER& RSF & Weibull\\
\hline
 $N$ & $10^4$ & $10^4$  &  $10^4$  &  $10^4$\\
 $|E|$ &  $10^5$  &  $5. 10^5$ & $22000$ & $55000$ \\
 $\overline{k}$  & $20$ & $100$  &  $4.4$  & $11$  \\
 $k_{max}$  &   $40$  &  $140$   &  $3500$  &  $2000$\\
\hline
\end{tabular}
\end{center}
\label{table}
\end{table}

\subsection{Betweenness centrality}

\begin{figure}[t]
\begin{center}
\includegraphics[width=8.0cm]{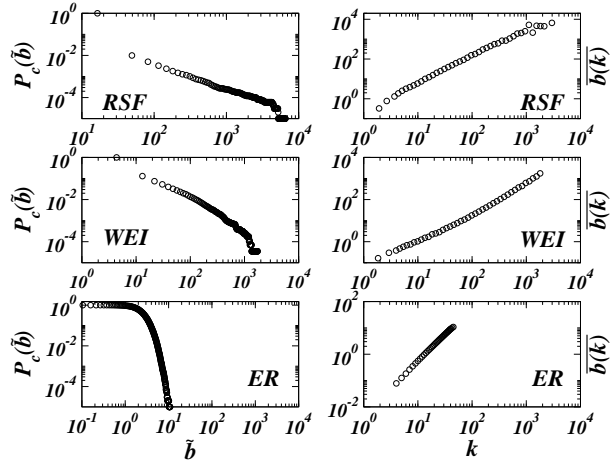}
\end{center}
\caption{Cumulative distribution of the rescaled vertex betweenness
  (left) and average behavior as a function of the connectivity (right) 
  in the graph models.}
\label{fig:2}
\end{figure}

Since the topological properties governing the traceroute exploration
is the betweenness centrality, it is worth reviewing its general
properties in the case of the models considered here. In Fig.~\ref{fig:2} we
report the vertex betweenness cumulative distributions for the ER model as
well as for the graphs with scale-free or Weibull distributions of
connectivity.

In homogeneous networks, the vertex and edge betweennesses are as well
homogeneous quantities and their distributions are peaked
around their average values $\overline{b}$ and $\overline{b_{e}}$,
respectively, spanning only a small range of variations. These values can thus
be considered as typical values. Moreover, the betweenness is correlated
with the degree, as shown by the study of the rescaled betweenness averaged
over vertices of given degree $k$, $\overline{b(k)}$, which increases with $k$.

For heterogeneous models, the betweenness distribution is
heavy-tailed, allowing for an appreciable fraction of vertices and edges with
very high betweenness\cite{bart}. In particular,
in scale-free graphs the site betweenness is related
to the vertices degree as $\overline{b(k)}\sim k ^\beta$, where $\beta$ is an
exponent depending on the model \cite{bart}. Since in heavy-tailed degree
distributions the allowed degree is varying over several orders of magnitude,
the same occurs for the betweenness values, as shown in Fig.~\ref{fig:2}, and
the tail of the distribution is broader the broader the connectivity
distribution: larger values are consequently reached for the RSF case with
$\gamma=2.3$ than for the Weibull case.

\section{Efficiency in the sampling of graphs}\label{efficiency}

\begin{figure}[t]
\begin{center}
\includegraphics[width=8.0cm]{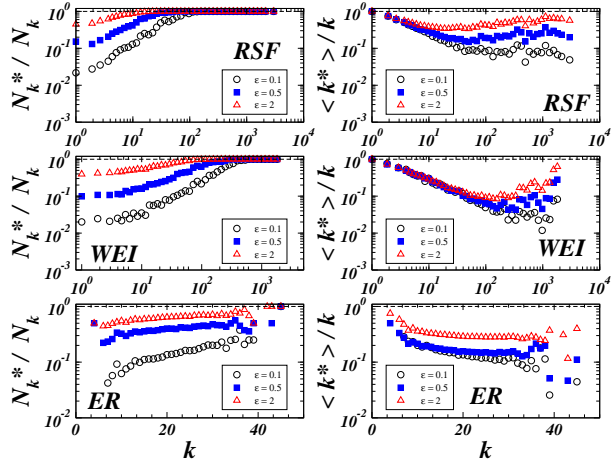}
\end{center}
\caption{Frequency $N_k^*/N_k$ of detecting a
vertex of degree $k$ (left) and proportion of discovered edges 
$\left\langle k^*\right\rangle/k$ (right) as a function of the degree
in the RSF, WEI, and ER graph models. The exploration setup considers $N_S=2$
and increasing probing level $\epsilon$ obtained by progressively
higher density of targets $\rho_T$. The axis of ordinates is in log scale to
allow a finer resolution.}
\label{fig:3}
\end{figure}

Let us first consider the case of homogeneous graphs.  Since a large majority
of vertices and edges will have a betweenness very close to the average value,
we can use Eq.~(\ref{edgesample}) and (\ref{vertexsample}) to estimate the
order of magnitude of probes that allows a fair sampling of the graph. Indeed,
both $\langle\pi_{i,j}\rangle$ and $\langle\pi_{i}\rangle$ tend to $1$ if
$\epsilon\gg$ max$\left[\overline{b}^{-1}, \overline{b_{e}}^{-1}\right]$. In
this limit all edges and vertices will have probability to be discovered very
close to one.
At lower value of $\epsilon$, obtained by varying $\rho_T$ and $N_S$, the
underlying graph is only partially discovered. Fig.~\ref{fig:3} shows 
the behavior
of the fraction $N_k^*/N_k$ of discovered vertices of degree $k$, where $N_k$
is the total number of vertices of degree $k$ in the underlying graph, and the
fraction of discovered edges $\left\langle k^*\right\rangle/k$ in vertices of
degree $k$. $N_k^*/N_k$
naturally increases with the density of targets and sources, and it
is slightly increasing with $k$. The latter behavior can be easily
understood by noticing that vertices with larger degree have on average a
larger betweenness. On the other hand, the range of variation of
$k$ in homogeneous graphs is very narrow and only a large level of probing
may guarantee very large discovery probabilities.  Similarly the behavior of
the effective discovered degree can be understood by looking at
Eq.~(\ref{degdisc}). 
Indeed the initial decrease of
$\left\langle k^*\right\rangle/k$ is finally compensated by the increase of
$\overline{b(k)}$.

The situation is different in graphs with heavy-tailed connectivity
distributions, for which the betweenness spans various orders
of magnitude. In such a
situation, even in the case of small $\epsilon$, vertices whose betweenness is
large enough ($b_i \epsilon\gg 1$) have $\left\langle
  \pi_i\right\rangle\simeq 1$. Therefore all vertices with degree $k\gg
\epsilon^{-1/\beta}$ will be detected with probability one. This is clearly
visible in Fig.~\ref{fig:3} where the discovery probability $N_k^*/N_k$ of
vertices with degree $k$ saturates to one for large degree values.
Consistently, the degree value at which the curve saturates decreases with
increasing $\epsilon$.  A similar effect is appearing in the measurements
concerning $\left\langle k^*\right\rangle/k$. After an initial decay 
(Fig.~\ref{fig:3}) the effective discovered degree is increasing with the
degree of the vertices. This qualitative feature is captured by
Eq.~(\ref{degdisc}) that gives $\left\langle k^*\right\rangle/k\simeq \epsilon
k^{-1}(1 + \overline{b(k)})$.  At large $k$ the term
$k^{-1}\overline{b(k)}\sim k^{\beta -1}$ takes over and the effective
discovered degree approaches the real degree $k$. Moreover, it appears
clearly that the broader the distribution of betweennesses or connectivities,
the better the sampling obtained.

\section{Redundancy and dissymmetry of the discovery process}\label{redundancy}

In this section we introduce tools suitable to estimate how
\texttt{traceroute}-like procedures discover the vertices and the
edges of the unknown underlying network.  
The most common biases affecting the mapping process concern the miss of
lateral connectivity, 
and the multiple sampling of central vertices (and edges), which
may affect the efficiency of the whole process. While the first problem
might be solved by an optimization in the deployment of probes, actually
relying on a criterion of decentralization of sources and targets,
multiple sampling can be studied through some general concepts like
the {\em redundancy} and {\em dissymmetry} of the discovery process.

\subsection{Redundancy}
  
Let us define the edge redundancy $r_{e}(i,j)$ of an edge $(i,j)$ in a
\texttt{traceroute}-sampling 
as the number of probes passing through the edge $(i,j)$.
Using the notations of section \ref{stat}, this quantity is written
for a given set of probes and targets as 
\begin{equation}
r_{e}(i,j) = \sum_{l \neq m} 
\left( \sum_{s=1}^{N_{S}} \delta_{l,i_{s}}
 \sum_{t=1}^{N_{T}} \delta_{m,i_{t}} \sigma_{i,j}^{(l,m)} \right) .
\label{edgeredund}
\end{equation}
Averaging over all possible realizations and assuming the uncorrelation 
hypothesis, we obtain
\begin{equation}
\left\langle r_{e}(i,j) \right\rangle = 
\simeq \sum_{l\neq m} \rho_T\rho_S \left\langle \sigma_{i,j}^{(l,m)}
\right\rangle = \rho_T\rho_S b_{ij} \ . \ \ \ \ \ \ 
\label{averedgeredun}
\end{equation}

This result implies that the average redundancy of an edge is related to the
density of sources and targets, but also to the edge betweenness. For example,
an edge of minimum betweenness $b_{ij} = 2$ can be discovered at most twice in
the extreme limit of an all-to-all probing. On the contrary, a very central
edge of betweenness $b_{ij}$ close to the maximum $N (N-1)$, would be
discovered with a redundancy close to $(N-1)$ by a \texttt{traceroute}-probing
from a single source to all the possible destinations.

Similarly, the
redundancy $r_{n}(i)$ of a vertex $i$, intended as the number of times the
probes cross the vertex $i$, can be obtained:
\begin{equation}
r_{n}(i) =  \sum_{l \neq m }
\sigma_{i}^{(l,m)} \sum_{s=1}^{N_{S}} \delta_{l,i_{s}} 
\sum_{t=1}^{N_{T}} \delta_{m,i_{t}} \ .
\end{equation}
After separating the cases $l=i$ and $m=i$ in the sum, the averaging over
the positions of sources and targets yields in the 
mean-field approximation:
\begin{equation}
\left\langle r_{n}(i)\right\rangle  =  \sum_{l \neq m \neq i}  \rho_{S}\rho_{T}
 \langle \sigma_{i}^{(l,m)} \rangle + 
2 \rho_{S} \rho_{T} N  \simeq  2\epsilon + \rho_{S}\rho_{T} b_{i} \ .
\label{avernoderedun}
\end{equation}
In this case, a term related to the number of traceroute probes $\epsilon$
appears, showing that a part of the mapping effort
unavoidably ends up in 
generating vertex detection redundancy.

\begin{figure}[t]
\begin{center}
\includegraphics[width=6.cm]{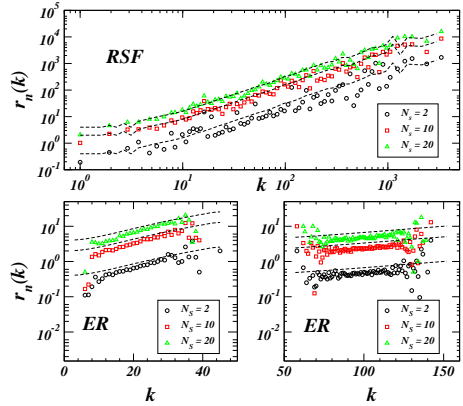}
\end{center}
\caption{
  Average vertex redundancy as a function of the 
  degree $k$ for RSF (top) and ER
  (bottom) model ($N = 10^{4}$). For the ER model, two blocks of data are
  plotted, for $\overline{k}=20$ (left) and for $\overline{k}=100$ (right) The
  target density is fixed ($\rho_{T}=0.1$), and $N_S=2$ (circles), $10$
  (squares), $20$ (triangles). The dashed lines represent the analytical
  prediction $2\epsilon + \rho_{S} \rho_{T} \overline{b(k)}$ in perfect
  agreement with the simulations.}
\label{fig:4}
\end{figure}
In Fig.~\ref{fig:4} we report the behavior of the average vertex 
redundancy as a
function of the degree $k$ for both homogeneous and heterogeneous graphs.  For
both models, the behaviors are in good agreement with the mean-field
prediction, showing the tight relation between redundancy
and betweenness centrality.

In the case of heavy-tailed underlying networks, the vertex 
redundancy typically
grows as a power-law of the degree, while the values for random graphs vary on
a smaller scale. This behavior points out that the intrinsic hierarchical
structure of scale-free networks plays a fundamental role even in the process
of path routing, resulting in a huge number of probes iteratively passing
through the same set of few hubs.  On the other hand, for homogeneous graphs
the total number of vertex discoveries is quite uniformly 
distributed on the whole
range of connectivity, independently of the relative importance of the vertices.

\subsection{Dissymmetry: Participation Ratio}

The high rate of redundancy intrinsic to the exploration process, however,
does not imply that the local topology close to a vertex is well discovered:
preferential paths could indeed carry most of the probing effort leading to
just a partial discovery of the vertex connections. This amounts to a
dissymmetry of the exploration process that probes some edges much more than
others, eventually ignoring some of those, in the neighborhood of a given
vertex.  Together with redundancy measures, let us consider the relative
number of occurrences of a given edge $(i,j)$ during the traceroute, with
respect to the total occurrence for the edges in the neighborhood of $i$.  For
each discovered vertex $i$, we can thus define a set of frequencies
$\{f_{j}^{(i)}\}_{j \in \mathcal{V}(i)}$ for the edges $(i,j)$ of its
neighborhood. In terms of redundancy the \textit{edge frequency} $f_{j}^{(i)}$
is defined as
\begin{equation} 
f_{j}^{(i)} = \frac{r_{e}(i,j)}{\sum_{j \in \mathcal{V}(i)} r_{e}(i,j)}, 
\ \ \ \  0 \leq f_{j}^{(i)} \leq 1 \ \ \forall (i,j) \in E, 
\end{equation}         
and indicates the probability that any given probing path discovering 
the vertex $i$, is passing by the edge $(i,j)$.
\begin{figure}[t]
\begin{center}
\includegraphics[width=7.cm]{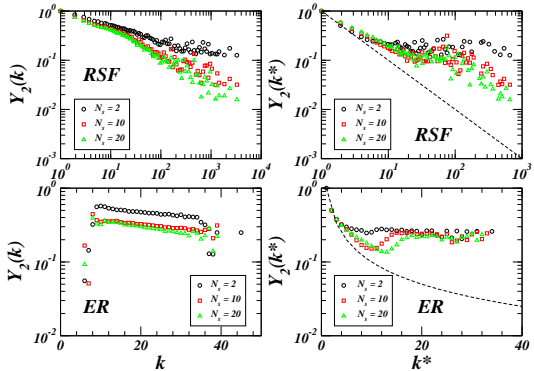}
\end{center}
\caption{
Participation Ratio as a function of real ($k$) and discovered ($k^*$)
  connectivity for RSF (top) and ER (bottom) models ($N=10^{4}$). The target
  density is fixed ($\rho_{T}=0.1$) and three value of $N_S$ are presented: 2
  (circles), 10 (squares), 20 (triangles).  The dashed lines correspond to
  the $1/k^*$ bound.}
\label{fig:5}
\end{figure}
The dissymmetry of the discovery of the neighborhood of a vertex may be
quantified through the \textit{participation ratio} of these frequencies:
\begin{equation}
Y_{2}(i) = \sum_{j \in \mathcal{V}(i)} {\left(f_{j}^{(i)} \right)}^2 \ .
\end{equation}   
If all the edge frequencies of $i$ are of the same order $\sim 1/{k^{*}_{i}}$
(only discovered links give a finite contribution), the participation ratio
should decrease as $1/{k^{*}_{i}}$ with increasing discovered connectivity
${k^{*}_{i}}$.  Hence, in the limit of an optimally symmetric sampling, it
should yield a strict power law behavior $Y_{2}(k^{*}) \sim {k^{*}}^{-1}$.  On
the contrary, when only few links are preferred, for instance because more
central in the shortest path routing, the sum is dominated by these terms,
leading to a value closer to the upper bound $1$.  Numerical data for $Y_{2}$
as a function of the actual ($k$) and discovered ($k^{*}$) connectivity for
different probing efforts, are displayed in Fig.~\ref{fig:5}.  For
heterogeneous graphs, the values of $Y_{2}$ tend towards the curve
${k^{*}}^{-1}$ for increasing $\epsilon$.  In both cases this behavior is
better achieved at high degree values.  The tendency of high degree vertices
to be better sampled in a more symmetrical way is evident in the diagram for
$Y_{2}(k)$, where a crossover at large degrees appears.  On the contrary, in
the homogeneous case (ER), the figures show a general high level of dissymmetry
persistent at all degree values, only slightly dependent on the actual
connectivity and the probing effort.

\subsection{Dissymmetry: Entropy Measure}

In order to provide an alternative and in some cases more accurate study of
the dissymmetry of the exploration process, we introduce a more refined
frequency, $f_{kj}^{(i)}$ defined as the number of probes passing through the
{\em pair} $(k,i)-(i,j)$ of edges centered on the vertex $i$. This is the
probability of a probe to traverse a couple of edges with respect to the total
number of transits through any of the possible couples of edges in the
neighborhood of $i$.  This frequency takes fully into account the path
traversing each vertex and the dissymmetry of the flow.  By means of this
frequency, we define an entropy measure providing supplementary evidence of
the tight relation between local accuracy, dissymmetry of sampling and
topological characterization of graphs.  Indeed, a \texttt{traceroute}
discovering vertices crossing a larger variety of their links, and with
different paths, is expected to be more accurate (and likely efficient) than
the one always selecting the same path.  In the same spirit of the Shannon
entropy, which is a good indicator of disorder, we define the \textit{local
  traceroute entropy} of a vertex $i$ by
\begin{equation}
h_i = - \frac{1}{\log{({k^{*}}_i(k^*_i -1))}} 
\sum_{k \neq j \in \mathcal{V}(i)} f_{kj}^{(i)}
\log{f_{kj}^{(i)}},
\end{equation}
where $\log{{k^{*}_i}}$ is simply a normalization factor. This
quantity 
is bounded in the interval $0\leq h(i)\leq 1$. The case 
$h_i=1$ is reached when all the frequencies of probes spanning  
the edge couples of the vertex are equal. The case $H\simeq 0$
corresponds to a
dominating frequency in a specific edge couple. 
Also in this case it is possible to study the degree spectrum $H(k)$ 
of the entropy by measuring 
the average entropy on vertices with given degree $k$.

\begin{figure}[t]
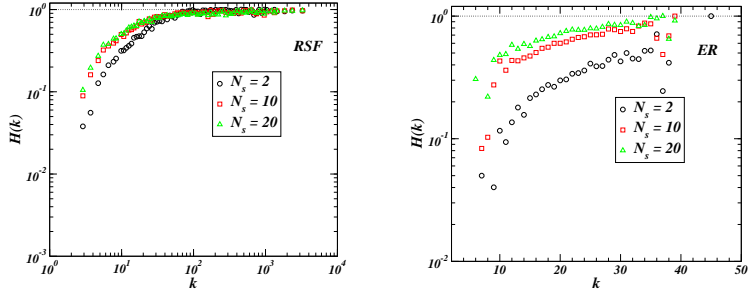

\begin{center}
\includegraphics[width=4.5cm]{entRSF}
\hskip .7cm
\includegraphics[width=4.5cm]{entER}
\end{center}
\caption{
 Entropy vs.~$k$: a saturation effect is clear at 
  medium-high degree vertices for scale free topologies (RSF), instead of a more
  regular increasing for homogeneous graphs (ER). In the figure there are
  different curves for $N_S$ = $2$ (circles), $10$ (squares), $20$ (triangles)
  and $\rho_{T} = 0.1$.}
\label{fig:6}
\end{figure}

The numerical data of $H(k)$ for RSF and ER models and for different levels of
probing are reported in Fig.~\ref{fig:6}. The values for ER are slightly
increasing both for increasing degree $k$ and number of sources $N_S$, with no
qualitative difference in the behavior at low or high degree regions.  On the
other hand, the case of heterogeneous networks agrees with the previous
observations. The curve for $H(k)$, indeed, shows a saturation
phenomenon to values very close to the maximum $1$ at large enough degree,
indicating a very symmetric sampling of these vertices.

In summary the previous studies indicates that in the case of
heterogeneous networks, the hubs and high betweenness vertices are in
general sampled redundantly, however, obtaining a rather symmetrical
discovery of their neighborhood. On the contrary, homogeneous networks
do not allow the presence of hubs and vertices are suffering a less
redundant sampling while showing a high dissymmetry of the local
exploration process. This results might be useful in deciding
source-target deployment strategies, by taking into account the
underlying topology of the network.   

\section{Degree distribution measurements}
\label{sec:pk}
\begin{figure}[t]
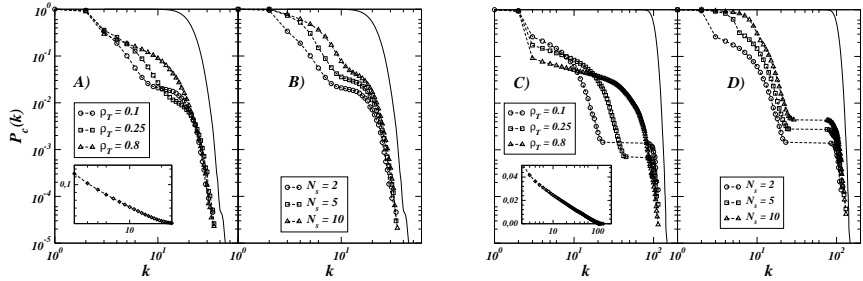

\begin{center}
\includegraphics[width=5.50cm]{PcumER}
\hskip .7cm
\includegraphics[width=5.0cm]{PcumER100}
\end{center}
\caption{
  Cumulative degree distribution of the sampled ER graph for USP probes.
  Figures A) and B) correspond to $\overline{k}=20$, and C) and D) to
  $\overline{k}=100$.  Figures A) and C) show sampled distributions obtained
  with $N_S=2$ and varying density target $\rho_T$. In the insets we report
  the peculiar case $N_S=1$ that provides an apparent power-law behavior with
  exponent $-1$ at all values of $\rho_T$, with a cut-off depending on
  $\overline{k}$.  The insets are in lin-log scale to show the logarithmic
  behavior of the corresponding cumulative distribution.  Figures B) and D)
  correspond to $\rho_T=0.1$ and varying number of sources $N_S$. The solid
  lines are the degree distributions of the underlying graph. For
  $\overline{k}=100$, the sampled cumulative distributions display plateaus
  corresponding to peaks in the degree distributions, induced by the sampling
  process.}
\label{fig:7}
\end{figure}

\begin{figure}[t]
\begin{center}
\includegraphics[width=7.cm]{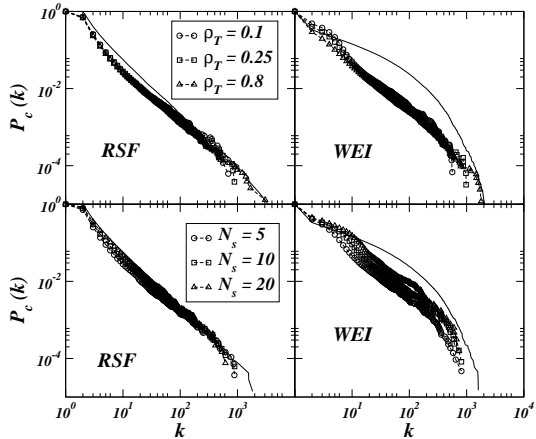}
\end{center}
\caption{Cumulative degree distributions of the sampled RSF and WEI graphs for
USP probes.  
The top figures show sampled distributions obtained with $N_S=5$ 
and varying density target $\rho_T$. 
The figures on the bottom correspond to
$\rho_T=0.25$ and varying number of sources $N_S$. The solid lines are the
degree distributions of the underlying graph.}
\label{fig:8}
\end{figure}

A very important quantity in the study of the statistical accuracy of the
sampled graph is the degree distribution. 
Fig.~\ref{fig:7} shows the cumulative degree distribution
$P_c(k^*>k)$ of the sampled graph defined by the ER model for increasing
density of targets and sources.  Sampled distributions are only approximating
the genuine distribution, however, for $N_S\geq 2$ they are far from true
heavy-tail distributions at any appreciable level of probing.  Indeed, the
distribution runs generally over a small range of degrees, with a cut-off that
sets in at the average degree $\overline{k}$ of the underlying graph. In order
to stretch the distribution range, homogeneous graphs with very large average
degree $\overline{k}$ must be considered; however, other distinctive spurious
effects appear in this case. In particular, since the best sampling occurs
around the high degree values, the distributions develop peaks that show in
the cumulative distribution as plateaus (see Fig.\ref{fig:8}). 
Finally, in the case of RSP and ASP model, we observe that the obtained
distributions are closer to the real one since they allow a larger number of
discoveries.

Only in the peculiar case of $N_S=1$ an apparent scale-free behavior with
slope $-1$ is observed for all target densities $\rho_T$, as analytically
shown by Clauset and Moore \cite{clauset}.  Also in this case, the
distribution cut-off is consistently determined by the average degree
$\overline{k}$.  It is worth noting that the experimental setup with a single
source is a limit case corresponding to a highly asymmetric probing process;
it is therefore badly, if at all, captured by our statistical analysis which
assumes homogeneous deployment.

The present analysis shows that in order to obtain a sampled graph with
apparent scale-free behavior on a degree range varying over $n$ orders of
magnitude we would need the very peculiar sampling of a homogeneous underlying
graph with an average degree $\overline{k}\simeq 10^n$; a rather unrealistic
situation in the Internet and many other information systems where $n\geq 2$.

In section \ref{efficiency}, we have shown clearly that, in
heterogeneous graphs, vertices with high degree are efficiently sampled with
an effective measured degree that is rather close to the real one. This means
that the degree distribution tail is fairly well sampled while deviations
should be expected at lower degree values.  This is indeed what we observe in
numerical experiments on graphs with heavy-tailed distributions (see
Fig.~\ref{fig:8}). Despite both underlying graphs have a small average
degree, the observed degree distribution spans more than two orders of
magnitude. The distribution tail is fairly reproduced even at rather small
values of $\epsilon$. The data shows clearly that the low degree regime is
instead under-sampled. This undersampling can either yield an 
apparent change in the exponent of the degree distribution (as also
noticed in \cite{delos} for single source experiments), or, if $N_S$ is
small, yield a power-law like distribution for an underlying Weibull
distribution. Furthermore,  as Fig.~\ref{fig:8} shows, an increase
in the number of sources starts to discriminate between scale-free and
Weibull distributions by detecting a curvature in the second case 
even at small values $\rho_T= 0.25$. It is, however, fair to say that
while the experiments clearly points out a broad and heavy-tailed
distribution, the distinction between different types of heavy-tailed
distribution needs an adequate level of probing.

In conclusion, graphs with heavy-tailed degree
distribution allow a better qualitative representation of their
statistical features in sampling experiments. Indeed, the most
important properties of these graphs are related to the heavy-tail
part of the statistical distributions that are indeed well
discriminated by the \texttt{traceroute}-like exploration. 
On the other hand, the accurate identification of 
the distribution forms requires a fair level of sampling that 
it is not clear how to determine quantitatively in the case of an unknown
underlying network. We will discuss the implications of these
results in real Internet measurements in Sec.~\ref{sec:conc}.

\section{Optimization of mapping strategies}
\label{sec:opt}
In the previous sections we have shown that it is possible to have a
general qualitative understanding of the efficiency of network
exploration and the induced biases on the statistical properties. 
The quantitative analysis of the sampling strategies, however, is a
much harder task that calls for a detailed study of the discovered 
proportion of  the underlying graph and the precise deployment of
sources and targets. In this perspective, very important quantities
are the fraction $N^*/N$ and $E^*/E$ of vertices and edges discovered
in the sampled graph, respectively. Unfortunately, the mean-field
approximation breaks down when we aim at a quantitative representation
of the results. The neglected correlations are in fact very important
for the precise estimate of the various quantities of interest. 
For this reason we performed an extensive set of numerical
explorations aimed at a fine determination of the level of sampling
achieved for different experimental setups.

\begin{figure}[t]
\begin{center}
\includegraphics[width=7.0cm]{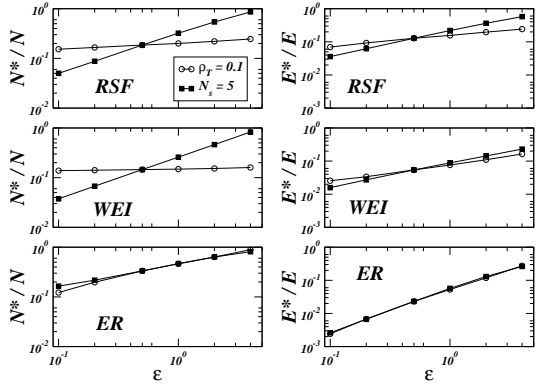}
\end{center}
\caption{Behavior of the fraction of discovered edges in explorations 
with increasing $\epsilon$. For each underlying graph studied we 
report two curves corresponding to larger $\epsilon$ achieved by increasing
the target density $\rho_T$ at constant $N_S=5$ (squares)
or the number of sources $N_S$
at constant $\rho_T=0.1$ (circles).}
\label{fig:14}
\end{figure}

In Fig.~\ref{fig:14} we report the proportion of discovered edges in
the numerical exploration of the graph models defined previously for
increasing level of probing $\epsilon$. The level of probing is
increased either by raising the number of  sources at fixed target density
or by raising the target density at fixed number of sources.
As expected, both strategies  are progressively more efficient with
increasing levels of probing.  
In heterogeneous graphs, it is also possible to see that when the 
number of sources is $N_S\sim
\mathcal{O}(1)$ the increase of the number of targets achieves better
sampling than increasing the deployed sources. On the other hand, it
is easy to perceive that the shortest path route mapping is a
symmetric process if we exchange sources with targets. This is
confirmed by numerical experiments in which we use a very large
number of sources and a number of targets  $\rho_T\sim
\mathcal{O}(1/N)$, where the trends are opposite: the increase of 
the number of sources achieves better sampling than increasing the 
deployed targets.

\begin{figure}[t]
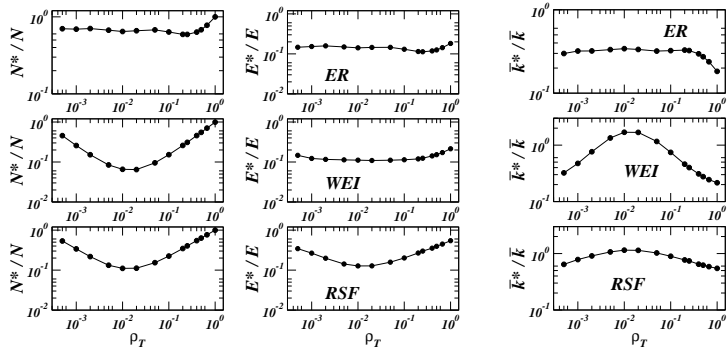

\vskip .5cm
\begin{center}
\includegraphics[width=6.0cm]{Nstar_Estar}
\hskip .5cm
\includegraphics[width=2.90cm]{kstar_cstar}
\end{center}
\caption{Behavior as a function of $\rho_T$ 
  of the fraction of discovered edges and vertices in explorations with fixed
  $\epsilon$ (here $\epsilon=2$).  Since $\epsilon=\rho_T N_S$, the increase
  of $\rho_T$ corresponds to a lowering of the number of sources $N_S$.
   The plots on the right show
the fraction of the normalized average degree $\overline{k}^*/\overline{k}$.}
\label{fig:15}
\end{figure}

This finding hints toward a behavior that is determined by the
number of sources and targets, $N_S$ and $N_T$. Any quantity is 
thus a function of $N_S$ and $N_T$,
or equivalently of $N_S$ and $\rho_T$. This point is clearly
illustrated in Fig.~\ref{fig:15}, where we report the behavior of
$E^*/E$ and $N^*/N$ at fixed $\epsilon$ and varying $N_S$ and $\rho_T$. 
The curves exhibit a non-trivial behavior and since we will work at
fixed $\epsilon=\rho_T N_S$, any measured quantity can then be written as
$f(\rho_T,\epsilon/\rho_T)=g_\epsilon(\rho_T)$.
Very interestingly, the curves show a structure allowing for local
minima  and maxima in the discovered portion of the underlying graph.

This feature can be explained by a simple symmetry argument. 
The model for \texttt{traceroute} is symmetric by the exchange of sources
and targets, which are the endpoints of shortest paths: an exploration with
$(N_T,N_S)=(N_1,N_2)$ is equivalent to one with  $(N_T,N_S)=(N_2,N_1)$.
In other words, at fixed $\epsilon=N_1 N_2/N$, a density of targets
$\rho_T=N_1/N$ is equivalent to a density
$\rho'_T= N_2/N$. Since $N_2=\epsilon/\rho_T$ we obtain that at
constant $\epsilon$, experiments with $\rho_T$ 
and $\rho'_T=\epsilon/(N\rho_T)$ are equivalent obtaining by symmetry
that any measured quantity obeys the equality 
$g_\epsilon(\rho_T)= g_\epsilon \left( \frac{\epsilon}{N\rho_T} \right)$.
This relation implies a symmetry point signaling  the presence of a
maximum or a minimum at $\rho_T =\epsilon/(N\rho_T)$. We therefore 
expect the occurrence of a symmetry in the graphs of 
Fig.\ref{fig:15} at $\rho_T\simeq \sqrt{\epsilon/N}$. Indeed, 
the symmetry point is clearly
visible and in quantitative good agreement with the previous
estimate in the case of heterogeneous graphs. On the contrary,
homogeneous underlying topology have a smooth behavior that makes
difficult the clear identification of the symmetry point. 
Moreover, USP probes create a certain level of
correlations in the exploration that tends to  hide the complete
symmetry of the curves.

\begin{figure}[t]
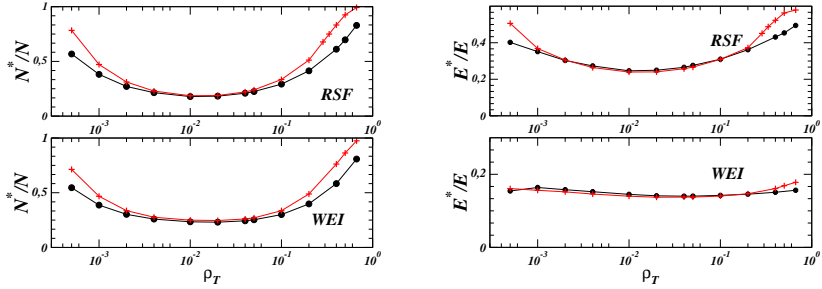

\begin{center}
\includegraphics[width=5.0cm]{lowBCNstarNx4eps2}
\hskip .7cm
\includegraphics[width=5.0cm]{lowBCEstarEx4eps2}
\end{center}
\caption{Behavior as a function of $\rho_T$  
  of the fraction of discovered edges and vertices in explorations with fixed
  $\epsilon$ (here $\epsilon=2$). The circles correspond to a random
deployment of sources and targets while the crosses are obtained when sources
and targets are vertices with lowest betweenness vertices. }
\label{fig:lowBC}
\end{figure}

The previous results imply that at fixed levels of probing $\epsilon$
different proportions of sources and targets may achieve different levels of
sampling. This hints to the search for optimal strategies in the relative
deployment of sources and targets.  The picture, however, is more complicate
if we look at other quantities in the sampled graph. In Fig.\ref{fig:15} we
show the behavior at fixed $\epsilon$ of the average degree $\overline{k}^*$
measured in sampled graphs normalized by the actual average degree
$\overline{k}$ of the underlying graph as a function of $\rho_T$. The plot
shows also in this case a symmetric structure.  By comparing the data of
Fig.\ref{fig:15} we notice that the symmetry point is of a different nature
for different quantities: the minimum in the fraction of discovered edges
corresponds to the best estimate of the average degree.  In other words, the
best level of sampling is achieved at particular values of $\epsilon$ and
$N_S$ that are conflicting with the best sampling of other quantities.

The evidence purported in this section hints to a possible optimization of the
sampling strategy. The optimal solution, however, appears as a trade-off
strategy between the different level of efficiency achieved in competing
ranges of the experimental setup. In this respect, a detailed and quantitative
investigation of the various quantities of interest in different experimental
setups is needed in order to pinpoint the most efficient deployment of
source-target pairs depending on the underlying graph topology.
While such a detailed analysis lies beyond the scope of the present study,
an interesting hint comes from the analytical results of section
\ref{stat}: since vertices with large betweenness have typically a very
large probability of being discovered, placing the sources and
targets preferentially on low-betweenness vertices (the most 
difficult to discover) may have an impact on the whole process. This
is what we investigate in Fig.~\ref{fig:lowBC} in which we report
the fraction of vertices and edges discovered by either
a random deployment of sources and targets or a deployment on
the lowest-betweenness vertices. It is apparent that such a deployment
allows to discover larger parts of the network. Of course the 
procedure used is unrealistic since identifying low-betweenness 
vertices is not an easy task. The usual correlation between connectivity
and betweenness however indicates that the exploration of a 
real network could be improved by a massive deployment of sources 
using low-connectivity vertices.

\section{Conclusions and outlook}
\label{sec:conc}
The rationalization of the exploration biases at the statistical 
level provides a general interpretative framework for the results 
obtained from the numerical experiments on graph models. The sampled
graph clearly distinguishes the two situations defined by homogeneous and
heavy-tailed topologies, respectively. This is due to the exploration process
that statistically focuses on high betweenness vertices, thus providing a very
accurate sampling of the distribution tail. In graphs with heavy-tails,
such as scale-free networks, the main topological features are
therefore easily discriminated since the relevant statistical
information is encapsulated in the degree distribution tail 
which is fairly well captured. Quite surprisingly, the sampling
of homogeneous graphs appears more cumbersome than those of
heavy-tailed graphs. Dramatic effects such as the existence 
of apparent power-laws, however, are found only in very peculiar cases. In
general, exploration strategies provide sampled distributions with 
enough signatures to distinguish at the statistical level between graphs 
with different topologies.
   
This evidence might be relevant in the discussion of real data from
Internet mapping projects. Indeed, data available so far
indicate the presence of heavy-tailed degree distribution both at the
router and AS level. In the light of the present discussion, it is
very unlikely that this feature is just an artifact of the mapping
strategies. The upper degree cut-off at the router and AS level 
runs up to $10^2$ and $10^3$, respectively. A homogeneous graph should
have an average degree comparable to the measured cut-off and this is
hardly conceivable in a realistic perspective (for instance, it would 
require that nine routers over ten would have more than 100 links to 
other routers). In addition, the major part of mapping projects are
multi-source, a feature that we have shown to readily wash out the  
presence of spurious power-law behavior. On the contrary,
heterogeneous networks with heavy-tailed degree distributions
are sampled with particular accuracy for the large degree
part, generally at all probing levels.
This makes very plausible, and
a natural consequence, that the heavy-tail behavior observed in 
real mapping  experiments is a genuine feature of the Internet.
Furthermore, heterogeneous graphs show a striking tendency to improve the 
mapping efficiency at large degree vertices, while exponential graphs 
seem to respond in a homogeneous way independent of the degree value.

On the other hand, it is important to stress that while at the qualitative
level the sampled graphs allow a  discrimination of the statistical
properties, at the quantitative level they might exhibit considerable
deviations from the true values such as size, average degree, 
and the precise analytic form of the heavy-tailed degree
distribution. For instance, the exponent of the power-law behavior
appears to suffer from noticeable biases. In this respect, it is of major
importance to define strategies that optimize the estimate of the various
parameters and quantities of the underlying graph. In this paper we have shown
that the proportion of sources and targets may have an impact on the accuracy
of the measurements even if the number of total probes imposed to the system
is the same. For instance, the deployment of a highly distributed
infrastructure of sources probing a limited number of targets may result as
efficient as few very powerful sources probing a large fraction of the
addressable space~\cite{tr@home}. 
The optimization of large network sampling is therefore an
open problem that calls for further work aimed at a more quantitative
assessment of the mapping strategies both on the analytic and numerical side.

\section*{Acknowledgments}

We are grateful to M. Crovella, P. De Los Rios, T. Erlebach, T. Friedman,
M. Latapy and T. Petermann for  very useful discussions and
comments.  This work has been partially supported by 
the European Commission Fet-Open
project COSIN IST-2001-33555 and contract 001907 (DELIS).

\end{document}